\documentclass[a4paper,UKenglish,cleveref, autoref, thm-restate]{lipics-v2021}

\nolinenumbers
\usepackage{longtable}
\newcommand{\formal}[1]{{\ttfamily\small #1}}
\usepackage{amsmath,amssymb}
\usepackage{fvextra} %

\usepackage{todonotes}

\usepackage[utf8]{inputenc} %
\usepackage[T1]{fontenc}    %
\usepackage{url}            %
\usepackage{booktabs}       %
\usepackage{amsfonts}       %
\usepackage{nicefrac}       %
\usepackage{microtype}      %
\usepackage{xcolor}         %
\usepackage{comment}

\usepackage{booktabs}
\usepackage{subcaption}
\usepackage{graphicx} %
\usepackage{pgfplots} %
\pgfplotscreateplotcyclelist{rw}
{solid, mark repeat = 1, mark phase = 1, mark = *, black\\
        solid, mark repeat = 1, mark phase = 1, mark = square*, black\\}

\usepackage{tikz} %
\usetikzlibrary{shapes,arrows}
\usepackage{pgfplots}
\usepackage{float}
\usepackage{bussproofs}
\usepackage{xspace}
\usepackage{amsmath}
\usepackage{mathtools}
\usepackage{doi}
\usepackage{multicol}

\usepackage{mathpartir}

\hypersetup{
  colorlinks=true,
  linkcolor=blue,
  urlcolor=blue,
  citecolor=blue
}

\usepackage{color}

\usetikzlibrary{arrows,shapes,calc}
\usetikzlibrary{trees,positioning,fit}
\tikzstyle{arrow}=[draw,-to,thick]
\tikzstyle{embedding} = [draw, minimum width=8mm, minimum height=6mm]
\tikzstyle{nnop} = [draw, minimum width=8mm, minimum height=8mm, rounded
corners]
\tikzstyle{block} =
[rectangle, draw,
minimum width=9mm, text centered, rounded corners, minimum height=2.0em]
\tikzstyle{smallblock} =
[rectangle, draw,
minimum width=9mm, text centered, minimum height=2.0em]

\tikzstyle{line}=[draw]
\tikzstyle{cloud} =
[draw, text centered, ellipse, minimum height=2.0em, minimum width=9em]

\usepackage[]{listings}

\newcommand{\mathleft}{\@fleqntrue\@mathmargin0pt}

\newenvironment{mypar}[2]
  {\begin{list}{}%
    {\setlength\leftmargin{4mm}
    \setlength\rightmargin{4mm}}
    \item[]}
  {\end{list}}

\title{130k Lines of Formal Topology in Two Weeks:\\Simple and Cheap Autoformalization for Everyone?}

\author{Josef Urban}{AI4REASON and University of Gothenburg}{josef.urban@gmail.com}{}{}

\authorrunning{J. Urban} %
\titlerunning{130k Lines of Formal Topology in Two Weeks} %

\Copyright{Josef Urban} %



\category{} %

\relatedversion{} %

\funding{This work was supported by the ERC project NextReason,
Amazon Research Awards, the Czech MEYS under the ERC CZ project \emph{POSTMAN}
no.~LL1902, Czech Science Foundation grant no. 25-17929X, and the sponsors of the AI4REASON institute.\footnote{\url{https://ai4reason.eu/sponsors.html}}}

\begin{document}

\maketitle

  \begin{abstract}
    This is a brief description of a project that has already
    autoformalized a large portion of the general topology from the
    Munkres textbook (which has in total 241 pages in 7 chapters and
    39 sections). The project has been running since November 21, 2025
    and has as of January 4, 2026, produced 160k lines of formalized
    topology. Most of it (about 130k lines) have been done in two weeks,
    from December 22 to January 4, for an LLM subscription
    cost of about \$100. This includes a 3k-line proof of Urysohn's
    lemma, a 2k-line proof of Urysohn's Metrization theorem, over 10k-line proof of the Tietze extension theorem, and many
    more (in total over 1.5k lemmas/theorems).

    The approach is quite
    simple and cheap: build a long-running feedback loop between an
    LLM and a reasonably fast proof checker equipped with a core
    foundational library. The LLM is now instantiated as ChatGPT
    (mostly 5.2) or Claude Sonnet (4.5) run through the respective
    Codex or Claude Code command line interfaces. The proof checker is
    Chad Brown's higher-order set theory system Megalodon, and the
    core library is Brown's formalization of basic set theory and
    surreal numbers (including reals, etc). The rest is some
    prompt engineering and technical choices which we describe
    here. Based on the fast progress, low cost, virtually unknown ITP/library, and the simple setup
    available to everyone, we believe that (auto)formalization may
    become quite easy and ubiquitous in 2026, regardless of which proof assistant is used.

\end{abstract}

\section{Introduction}

\begin{mypar}{10mm}{5mm}
{\it
``It is the view of some of us that many people who could have easily contributed to project QED have been distracted away by the enticing lure of AI or AR.'' 

{\hfill\rm-- The QED Manifesto}
}
\end{mypar}

\paragraph*{Caveat:} This is an initial write-up about an ongoing
experiment (mostly to allow others to try it in their settings). My
assessment may evolve as the experiment proceeds.

\paragraph*{How it started:} I have been working (more or less seriously) with my colleagues on
learning-assisted autoformalization since 2014\footnote{I have likely
  coined the ``auto-formalization'' term in this context in my CICM'14
  talk \url{https://t.ly/PaeKH}~\cite{DBLP:conf/mkm/KaliszykUVG14}.}~\cite{DBLP:conf/mkm/KaliszykUVG14,DBLP:conf/itp/KaliszykUV15}. A bit of autoformalization history is given in my
2024 Bonn talk\footnote{\url{https://www.youtube.com/watch?v=4JeezEGc_gQ}} and its slides.\footnote{\url{http://grid01.ciirc.cvut.cz/~mptp/bonn24_af.pdf}} Long story
short, since the first MaLARea experiments in 2005/6~\cite{DBLP:conf/cade/Urban07}, I have become
quite convinced about the power of feedback loops between
reasoning/logic/deduction and learning, and believed that we are
ourselves in such a feedback loop when we formalize math (see e.g. the last two sentences in~\cite{DBLP:conf/lpar/Urban08}).
In 2018, I got further
impressed by the neural language models (especially attention-based)
showing surprisingly good informal-to-formal translation capability in
our first experiments over a large synthetic Mizar-to-LaTeX
corpus~\cite{DBLP:conf/mkm/WangKU18}. I started to believe my own
autoformalization propaganda, in the sense that autoformalization
would become usable much sooner than my 2014 conservative 25-year horizon
estimate.\footnote{\url{http://ai4reason.org/aichallenges.html}}
Since 2020, the field has become increasingly
popular in industry teams with much more resources, and (seeing their heavy buy-in\footnote{Sometimes perhaps too heavy - see the claims about ``introducing autoformalization'' at \url{https://archive.ph/8Ib1k}.}) I have (with some exceptions) mostly focused on other (less resource-heavy) feedback loops in ATP and conjecturing~\cite{DBLP:conf/mkm/UrbanJ20,DBLP:conf/itp/JakubuvCGKOP00U23,DBLP:journals/ijar/GauthierOU23,DBLP:conf/cade/GauthierU25,DBLP:journals/jsc/PiepenbrockUKOHJ25}.
I have occasionally tried to get ChatGPT (and similar LLMs)
to autoformalize smaller papers (such as the Ramsey(3,6) 4-page paper~\cite{cariolaro1999ramsey}) since it
appeared, but it initially didn't seem very good at carrying out
deeper and longer work -- at least not without building some infrastructure
around it.

The immediate impulse for starting the experiment described here were
Zar Goertzel's reports of good autoformalization progress with
coding agents such as Codex and Claude Code. In particular, in late
2025 (overlapping with this work) Zar managed to finish a full Lean proof of my testing paper on
Ramsey(3,6) with Claude Code in about 13k lines of Lean.\footnote{\url{https://github.com/zariuq/ai-agents/tree/main/lean-projects/ramsey36/Ramsey36}} After some
false starts and brief discussion with Chad Brown, I have decided to set up the current experiment in
formalization of topology with the Megalodon proof assistant and the
LLM-based command-line coding agents.

\section{Current Setup}

The latest setup has been (almost) stable since December 22 and used
for (almost) automated formalization resulting in about 130k lines
code and the above mentioned fully proved major theorems. The setup is as follows:

\begin{enumerate}
\item A ChatGPT Pro (\$200/month) subscription, giving access to the OpenAI's Codex coding agent.
\item I always use the ChatGPT 5.2 model with high (not extra high) reasoning.\footnote{Users with more resources could try to repeat the experiment with other settings and provide further ablations. I don't use the extra high setting because already the high reasoning depletes its weekly usage limit in less than a week.}
\item I isolate the coding agent using the Linux bubblewrap (bwrap\footnote{\url{https://github.com/containers/bubblewrap}}) sandboxing tool. I could have likely used any other reasonable sandboxing solution. I want to give the coding agents unlimited powers (calling all sorts of dangerous tools), hence their isolation inside the sandbox. I map several outside directories into the sandbox with various binaries and project data. While no serious harm was done by the agents so far inside the sandbox, there were already several occasions (like the agents getting confused about data location, file names, etc) that made me feel vindicated about having the isolation. I periodically back up the project data directory from the outside.
\item Inside bwrap, I run the command-line coding agent (now Codex, but I can and did run Claude Code too with similar approach and results).
\item Bwrap itself is run in the tmux terminal manager/multiplexer, typically in a long-running (multiday) session that is being logged and interacted with from the outside. I could likely have used other similar tools for this such as GNU Screen.
\item I use the Megalodon (higher-order set theory) proof checker binary for the proof checking. The binary used is the one from the Megalodon Wiki (mgwiki) Github repo, which ensures compatibility with the mgwiki toolchain. This allows to push any compiling   formalization into the mgwiki, where it is immediately html-ized. This cross-linked HTML presentation allows us to easily study the developing formalization.
\item As the
    core library I use Brown's formalization of basic set theory and
    surreal numbers (including reals, etc). We deleted most of the proofs there, replacing them with ``admit.'', so that the initial library file is smaller and (hopefully) easier to study for the LLMs. I however leave several of Brown's proofs there (such as Diaconescu proof of excluded middle), so that the LLM can (hopefully) study how to prove things in Megalodon.
  \item That's it. The rest is some prompting and some additional tools (optional and added later) described below.
\end{enumerate}

\section{Current Prompt}

With only a few exceptions, since December 22, the following short constant prompt has been repeatedly given to Codex each time it reports finishing its work (i.e., each time Codex CLI passes the control back to the user, asking for another prompt):\\\\
\texttt{Read the file CLAUDE.md . Treat it as authoritative work instructions. Follow those instructions exactly for all
  subsequent actions and responses. That means work as long as possible (as specified) without stopping.}\\
\\
My estimate is that this prompt has been automatically given to Codex about 1000-2000 times during the two weeks.\footnote{This can likely be calculated precisely after parsing the Codex logs.}
The current (14th) version of the ``rules of work'' file CLAUDE.md is as follows:
\begin{small}
\begin{verbatim}
# Rules for Working on Math_Background.mg

## STRICT PROHIBITIONS

### **ONLY edit in the topology section**
- The topology section starts at **line 6495** (`Section Topology.`)
- Everything after line 6495 is by default in the Topology section and can be
  edited (even very high line numbers).
- **NEVER edit anything before line 6495**
- This includes set theory, cardinal arithmetic, ordinals, etc.

### Before ANY edit to Math_Background.mg:
1. Verify the line number is >= 6495 (topology section)
2. If unsure, run: `grep -n "^Section Topology" Math_Background.mg`
3. If the target is before line 6495, **STOP** and ask the user

## When Resuming from Summary
1. **READ THIS FILE FIRST** before making any edits
2. Verify you understand which section you're allowed to edit
3. Double-check line numbers before proceeding

## Never throw away useful work
1. You should almost never revert to previous backups (you can use temporary
   admits when something is hard).
2. If there is a really major reason for reverting, you have to salvage all
   the useful work done in between.
3. In particular, you have to ensure that all compiling theorems and
   definitions added since are kept.
4. If you later discover such screwup, you have to immedially start working
   on salvaging the lost work.
5. This rule exist because you have acidentally thrown away 9k lines (more
   than a day) of work between bck2658 and bck2659 (by mysteriously justifying
   it as "reverting to colleague baseline"). This must never ever happen again.
6. **SIMPLE CHECK**: after each backup, run wc on the current and previous backup.
   If the current backup is smaller, you have to explicitly justify (in the
   CHANGES file) the decrease and explain that you have not thrown out useful work.

## Megalodon Language Details

### Logic System
- **This is classical Megalodon** (not constructive)
- Law of excluded middle (xm) is available
- Can use proof by contradiction freely

### Syntax Rules
- **`/\` is left-associative**: `A /\ B /\ C` means `((A /\ B) /\ C)`
- **Only use `(** … **)` for comments** - no other comment syntax; no `*` inside comments
- When working with nested conjunctions, remember the left-associative structure

## Work Strategy
- You can do the following things in any order but you should always
  progress and produce some more code.
- Keep carefully fixing any incorrect/bad definitions/theorems you
  find (note that this may lead to fixing some proofs, etc).
- Keep eliminating axioms and stubs, replacing them with more complete
  theorems and definitions (gradual/partial approaches are ok when needed).
- Keep removing admits in unfinished proofs and producing more complete
  proofs. This may also lead to adding more auxiliary lemmas/theorems.
- While doing the above, remember that:
- Doing easy things is initially OK, however, don't be afraid to try
  to do (gradually/partially if needed) some harder theorems too. Don't
  endlessly jump around looking for easy things - that wastes time.
- Balance simple infrastructure theorems with more challenging results
- Your **strong focus** should be on finishing the major well-known theorems
  ASAP (across all sections). Prioritize them (even if they are hard) over doing
  many examples/exercises.
- Use gradual/partial approaches for difficult theorems when needed (and don't
  delete such started partial proofs - use temporary "admit" in their various
  branches and keep gradually eliminating those). Also, structure bigger proofs
  into useful top-level/helper lemmas wherever possible.
- Frequently lookup the necessary notions defined before the topology section. Those
  should be completely trusted and built upon. They are listed in TRUSTED_DEFS.txt .
- Also, completely trust every theorem in the pre-topology section (and never
  attempt to re-prove it). We do have their proofs but have admited them in this
  document for brevity.
- Also, always grep all current Definitions and Theorems in the topology
  section before creating a new one. Be sure to remove/avoid duplicities.
- Do not introduce axioms that are exactly the same as the theorems you want to prove
  - it only duplicates and pollutes. In general, try to avoid adding axioms - it's
  better to have admitted theorems that will be (hopefully) eventually proved.
- Strongly prioritize properly finishing/debugging all proper definitions and
  (stating/proving) theorems from *all* sections of topology.tex . Doing the
  exercises is not as time-critical (although it may be occasionally needed as a
  prerequisite and generally useful, so use your judgement when to work on those admits).

### PROGRESS MONITORING
- Keep numbered PROGRESSXXXX files. For each section (i.e. starting with 12 and
  ending with 50) monitor the progress (stubs only, definitions/theorems properly
  stated, theorems fully proved, section mostly/fully completed, exercises
  mostly/fully completed). I.e., this PROGRESS file has to have a status
  line/paragraph for each of these 39 sections, which will be gradually changing
  towards more advanced levels.
- Make sure (and record in PROGRESSXXXX) that you make significant progress towards
  completing **ALL** (i.e. also later) sections. Dont get stuck only in the early
  sections. Use PROGRESSXXXX to find the imbalances and refocus on neglected parts
  (within means - dont jump around like crazy when working on something).

### Dependency/Admit Status/Monitoring
- after each numbered backup, run `mgdeps6.pl bckXXXX > DEPSXXXX`
- this will create DEPSXXXX file with one (topology section) theorem
  per line with info as follows `topology_elem_subset: lines:12,
  admit:NO, recadmit:NO, deps(2):[topology_on:D,topology_sub_Power:T].`
- this says if the theorem is (recursively) admitted and its direct topology
  deps (D for a def and T for theorem).  (note that the lines info there is
  after deleting all comments and that all non-topology deps are ignored)
- Use this to monitor the progress and find admitted theorems (major
  bottlenecks) that have to be proved so that others are recursively admit-free.

## Compilation Checking
- **Run megalodon frequently** to check for compilation errors
- **When megalodon produces no output, the code is correct**
- **Megalodon only prints output when there are errors or warnings**
- Silent output = success
- After any significant change, run: `timeout 30 megalodon -owned
  ownedSep132023 Math_Background.mg`
- Often do numbered backups like bck1107
- With each numbered backup, also write the numbered summary changes
  file like CHANGES1107 (it should really be a summary, not just a simple diff).
- You can lookup your previous work in these CHANGES files when unsure how to continue.
- Never overwrite an older backup file. The numbering has to continue from the
latest number. You must find it by running in tst7: ls bck* | sed 's/[^0-9]*//g'
| sort -n | tail -n 1.

## Why This Rule Exists
- Work before the topology section may have external dependencies
- Topology section is self-contained and safe to modify
- Breaking this rule wastes time undoing changes

## Last Updated
2025-12-02: Initial creation after accidentally editing lines 1728-1761
\end{verbatim}
\end{small}  

\subsection{Prompting Automation}
Before December 22, I was ``babysitting'' the command-line interface, i.e.,
looking every now and then if the agent has finished its work and waits for
another prompt. This was manageable, because my early \$20/month LLM
subscriptions would quickly run out of credits anyway and most of the time would
be spent waiting for the weekly credit limit to be reset.

With the \$200/month ChatGPT Pro subscription the limits on usage got much higher and automating the prompting became necessary. I have implemented it by writing the following script that watches the tmux session and feeds it the prompt followed by ``Enter'' whenever there has not been any change in the session for 60 seconds:

\begin{small}
\begin{verbatim}
#!/usr/bin/env bash

OUTDIR="/home/urban/ec/claude-code/tmux"
INTERVAL=60
mkdir -p "$OUTDIR"

while true; do
  ts=$(date +"%
  tmux capture-pane -p -S -1 > "$OUTDIR/capture_$ts.txt"
  # get two most recent captures
  mapfile -t files < <(ls -1t "$OUTDIR"/capture_*.txt 2>/dev/null | head -n 2)
  if [[ ${#files[@]} -eq 2 ]]; then
    if diff -q "${files[0]}" "${files[1]}" >/dev/null; then
      tmux send-keys 'Read the file CLAUDE.md . Treat it as authoritative work instructions.
Follow those instructions exactly for all subsequent actions and responses. That means work
as long as possible (as specified) without stopping.'; sleep 2; tmux send-keys Enter
    fi
  fi
  sleep "$INTERVAL"
done
\end{verbatim}
\end{small}  

\section{Other Prompts}

\subsection{Initial}
Initially, I have done quite a lot of experiments with the prompting, and also
had some special prompts for special tasks like the initial creation of all
definitions and theorems/lemmas corresponding to the textbook.
Here are the first five prompts from November asking Codex to do topology:\\

\begin{enumerate}
\item \texttt{Now that you have learned the basics we will do something more serious. First  read the file topology.tex . Then start formalizing it in Megalodon. The background from which should  learn the syntax of the definition, theorem and proofs and build on it is in the tst7/Math\_Background.mg . Here is  how you verify the article written so far: megalodon -owned ownedSep132023 Math\_Background.mg . Always make  numbered backups of the .mg files as before. Do 5 attempts and then stop.}
\item \texttt{Good, please continue. 5 more attempts.}
\item \texttt{Good. But maybe don't focus on the proofs so much and instead push forward to cover all the definitions and toplevel theorems/lemmas from the tex file. 5 more attempts.}
\item \texttt{Good. But maybe don't focus on the proofs so much and instead push forward to cover all the definitions and toplevel theorems/lemmas from the tex file. 5 more attempts. Try to do more without waiting for me, especially when things compile. Feel free to do the whole tex file at once.}
\item \texttt{Good. Now let's stop for a moment with the formalization. Instead, put into comments before each current definition and theorem the name, number, page number, location and/or statement from the original tex files, so that the reader can easily check that the from/informal items correspond to each other. Note that the comment syntax is as follows: (** from Part 1 **) .}
\end{enumerate}

A major initial effort was to make the coding agent work as independently as
possible and as long as possible, without them coming back, asking additional
questions, or being derailed in various ways. Ultimately, I started to get
independent runs of over 1 hour, and with the Pro subscription sometimes even 2
hours, before giving it the next prompt (which was very often just the same
prompt over and over, until I automated it fully as explained above).

\subsection{Focused Prompting}

On December 31, the system has already done quite some work on the Urysohn lemma but then started to switch to other things. Since this looked like an interesting end-of-year challenge, I issued the following special prompt:\\\\
\texttt{Special rule for
 today: Return to the proof of Urysohn lemma and finish it. Don't switch to other things before it is fully finished without any admits, axioms, etc. Finishing it is your singular goal for now.}\\\\
This has backfired a bit because the system then started to try all sorts of normally forbidden tricks. I told it to keep following the CLAUDE.md rules but focus on finishing Urysohn. The 3k-line proof indeed got finished by the end of 2025.

I made a similar focused attempt on January 4, after the system made a significant (but likely redundant -- see below) initial effort on the Tietze extension theorem, but then started to switch to other things. Since this was interesting I tried to focus it as follows (this time I was wiser and also told it to follow the rules - which worked well):\\\\
\texttt{Read the file CLAUDE.md . Treat it as authoritative work
  instructions. Follow those instructions exactly for all subsequent actions and
  responses. That means work as long as possible (as specified) without
  stopping. Special rule for today: Return to the proof of
  Tietze\_stepII\_real\_extension\_nonempty and finish it. Don't switch to other
  things
  before it is fully finished.}\\\\
This indeed led to the massive effort that finished the 10k-line proof (and all
its prerequisites) after about 20 hours. I believe this now provides quite a
usable way of combining the system's automated work on a large (textbook) formalization with
the user occasionally impatiently overriding it when he wants a particular proof
done first.

\section{Further Utilities}

\subsection{Hammer (aby) attempts}

Megalodon has since 2024 a simple higher-order hammer ``aby''~\cite{DBLP:conf/mkm/BrownKSU25}. In particular, it
can export the problems to the TH0 TPTP format and call e.g. higher-order
Vampire. There is so far no proof reconstruction (as e.g. in
Naproche). Initially it was tempting to save the LLM the work on detailed
Megalodon proofs and instead employ the hammer as much as possible. I have
written the scripts that automate that (to some extent) but so far this has  not
been very useful and likely requires further work on caching and (at least simple)
premise selection, possibly also some of the well-known translations to FOF
(first-order). A major issue that should be addressed (in the absence of proof reconstruction) is smart re-running of the ATPs when something changes.

\subsection{Theorem and Dependency Tracking}

Relatively late in the development I wanted the LLM to have a better overview of
the theorem dependencies and the biggest ``choke points''. It looked like for
example the completeness of reals was a major cause of recursive admits, even
though it should be quite feasible based on Brown's surreal numbers development
(which turned out to be true - see R\_lub\_exists below). So I have added a
simple dependency tracking script that at every iteration gets called on the
current file and produces for each theorem a one-line info with its current status
(fully proved, admitted, recursively admitted), length of the (partial) proof,
and the current proof dependencies. It seems that this has helped the LLM to
orient itself better (see also the advice about this in CLAUDE.md).

\section{Enforcing the order of work}
This is a relatively uncharted territory. People who formalize math
have developed good routines, but I am not aware of some ``handbook of
good formalization practices''.

Some of the problems that appeared were:

\begin{enumerate}
\item The LLMs were initially (occasionally) lazy in formalizing the theorem and definition statements. Sometimes they created stubs, ad-hoc or special cases instead of general cases, etc. This is dangerous if the LLM later forgets about it (its context window is finite) and takes such simplified specs seriously. I have countered it by telling it to be very cautious about such things, and once or twice even doing a special prompt to assess/remove the inaccuracies.
\item Initially it had a tendency to go for easy lemmas (often exercises/examples), rather than the hard/long theorems. Also, it could endlessly jump around searching for something easy. Again see CLAUDE.md for my attempts to prevent that (without being too strict).
\item I also wanted it to be continuously aware of all sections, rather than getting stuck in just one section for long time. Because this could later lead to major refactoring. I have tried to convince it to keep a PROGRESS (status) file (see CLAUDE.md) but this attempt has so far only mixed results. The trade-offs between going deep and proving hard theorems and being aware of the full formalization picture are obviously nontrivial also in large-scale human formalization projects.
\end{enumerate}

\section{Formalization Growth}

Table \ref{tab:lines} shows the number of lines after each 100 of backups (corresponding to commits). One can see the accelerated growth after switching to ChatGPT Pro.

\begin{table}[h]
\centering
\small
\setlength{\tabcolsep}{6pt}
\begin{tabular}{rr|rr|rr|rr}
\toprule
\multicolumn{8}{c}{\textbf{Commit number $\rightarrow$ number of lines}} \\
\midrule
\multicolumn{2}{c}{Commit \quad Lines} &
\multicolumn{2}{c}{Commit \quad Lines} &
\multicolumn{2}{c}{Commit \quad Lines} &
\multicolumn{2}{c}{Commit \quad Lines} \\
\cmidrule(lr){1-2}\cmidrule(lr){3-4}\cmidrule(lr){5-6}\cmidrule(lr){7-8}
100  & 1364   & 1000 & 7565   & 1900 & 44529  & 2800 & 78777  \\
200  & 2033   & 1100 & 9919   & 2000 & 47542  & 2900 & 83890  \\
300  & 3012   & 1200 & 11737  & 2100 & 52695  & 3000 & 90028  \\
400  & 3523   & 1300 & 12620  & 2200 & 56411  & 3100 & 96056  \\
500  & 3875   & 1400 & 13687  & 2300 & 56210  & 3200 & 105042 \\
600  & 4447   & 1500 & 19891  & 2400 & 62531  & 3300 & 116433 \\
700  & 4500   & 1600 & 31275  & 2500 & 68110  & 3400 & 119139 \\
800  & 4561   & 1700 & 31401  & 2600 & 72784  & 3500 & 128287 \\
900  & 4850   & 1800 & 39832  & 2700 & 68936  & 3600 & 137231 \\
     &        &      &        &      &        & 3700 & 148283 \\
\bottomrule
\end{tabular}
\caption{Growth of the formalization: commit number versus number of lines\label{tab:lines} }
\end{table}

\section{Resources Used}
While the Pro subscription is indeed only \$200/month, I have tried to use
independent tools to measure the USD cost of calling the LLM. In particular, I
call the \texttt{ccusage} tool every hour. Table~\ref{tab:costs} shows the reported costs for
three days of running, including a compactification related restart. Note that
these numbers are much higher than the \$100 spent over the two weeks. One
wonders what kind of business policies the LLM companies are implementing
(vicious fight for the market/subscribers?, overcharging companies for the
API usage?, trying to get as much data as possible?, bait-and-switch?, etc).

\begin{table}[h]
\centering
\small
\setlength{\tabcolsep}{4pt}
\begin{tabular}{rrrrrr}
\toprule
1--12 & 13--24 & 25--36 & 37--48 & 49--60 & 61--72 \\
\midrule
1918 & 2096 & 2229 & 2367 & 2549 & 602 \\
1932 & 2112 & 2234 & 2378 & 2565 & 618 \\
1949 & 2127 & 2239 & 2394 & 2576 & 635 \\
1964 & 2145 & 2243 & 2409 & 2576 & 651 \\
1980 & 2163 & 2254 & 2427 & 2576 & 664 \\
1993 & 2177 & 2268 & 2442 & 2576 & 678 \\
2008 & 2192 & 2282 & 2459 & 2576 & 694 \\
2024 & 2206 & 2296 & 2475 & 2576 & 707 \\
2040 & 2210 & 2312 & 2488 & 475  &      \\
2053 & 2215 & 2326 & 2503 & 489  &      \\
2067 & 2220 & 2339 & 2518 & 506  &      \\
2080 & 2224 & 2354 & 2534 & 519  &      \\
\bottomrule
\end{tabular}
\caption{Sequential hourly cost reports (hours 1--72), arranged in 6 columns of 12\label{tab:costs}}
\end{table}

\paragraph*{Running into Usage Limits}
In the semi-automated early mode, I managed to deplete the usage credits
for the \$20/month Codex or Claude Code subscription quite quickly. Still, on
these very cheap subscription the first 30k lines got produced in about a
month. This may be already very useful for many people.
I have been using the \$200/month Pro subscription for two weeks now and ran
into the usage limits every week. In the first week, the credits lasted for
almost the whole week, except for about half a day. In the second week, the weekly
credits got depleted earlier (on January 5, 3 days before the usage reset),
about 8 hours after finishing the ``battle of the Tietze Hill'' described
below. There are probably multiple factors influencing this.

\section{Miscellaneous Comments}

In general, I currently do not have the time and energy to do a full-scale
analysis of what went well and what went wrong. LLMs and other tools
could be used in the future to find automatically interesting cases from
the complete logs (which I am keeping) and the publicly available git commits.\footnote{\url{https://github.com/mgwiki/mgw_test/commits/main/}}
Perhaps this could lead to a
feedback loop augmenting the currently used prompts and tools based on such analysis.

\subsection{Context Compactification}
This is a major issue with the LLM-based technology I use. Every now and then, the
coding agent needs to get rid of unessential parts in its (limited) context
window and keep only the parts essential for further processing. The results of the
compactification (which is likely as proprietary as the LLMs I use) can be quite unpredictable.

Even worse, there seems to be a hard limit on how much compactification can be
done for a continuous session. I have reached this hard limit already twice:
once (December 26) when the ChatGPT session file reached about 252MB, and
another time (January 2) when the session file reached 351MB. Once this hard
limit is reached, it seems the session can no longer be continued in any way.
The standard advice I found seems to be ``do not run long sessions''. Instead,
people should maintain the essential status and progress themselves when doing
longer projects like these, always starting a new session with such updated status/progress.

Since this looked like a bigger project and I liked how the long-running session
behaved (not wanting to lose all it learned), I decided to ignore this
advice. Instead of resetting completely, I reset to an earlier version of the
session. In particular, when I reach the hard limit, I just rewrite the (big)
ChatGPT session (jsonl) file with a (much) younger (and smaller) version of
it. In particular, for both resets I used an early 65MB-big version of the
session file from December 18. This is likely an unsupported and dangerous
trick, but it has so far worked quite well for me.

\subsection{Running Megalodon}

As the file grows, running Megalodon gets slower and possibly problematic. While it took under 1s on the initial library file, the latest verification on my notebook takes 32s. However the file has  now almost 170k lines and 8MB:
\begin{small}
\begin{verbatim}
mptp@ar-2:~/ec/mizwrk/tst7$ time /home/mptp/ec/repos-mws/mgw_test/bin/megalodon bck4003
real	0m32.251s
user	0m32.095s
sys	0m0.154s
mptp@ar-2:~/ec/mizwrk/tst7$ wc bck4003 
 169122 1241687 7807911 bck4003
\end{verbatim}
\end{small}
An obvious solution is to start to split the file or to start admitting (the
\texttt{@proof} pragma in Mizar) the finished theorems. So far, I haven't done
that because: (i) I prefer the proofs to be possibly refactored by the LLMs,
(ii) running Megalodon doesn't seem to be the bottleneck compared to the LLM
usage and available credits. I.e., the number of calls to the LLM is still quite
high and even with the Pro subscription I run out of the LLM credits before the
usage limit resets.

\section{Major Theorems Proved}
Table~\ref{tab:thm300l} shows the proved theorems with a proof length of over 300 lines (no admits or recursive admits)
sorted according to the length of their direct Megalodon proof. I realize that this
may omit some major results which are stated as easier corollaries (see e.g. the
next section), however it still has some informative value. The theorem names in
the table are clickable and linked to their HTML presentation in mgwiki.

\newcommand{\mgw}[1]{\href{https://mgwiki.github.io/mgw_test/topology.mg.html\##1}{\texttt{\detokenize{#1}}}}

\begin{longtable}{p{0.72\linewidth}rr}
\caption{Topology theorems (mgwiki links), with line counts and dependency counts.\label{tab:thm300l}}\\
\toprule
Theorem & Lines & Deps \\
\midrule
\endfirsthead

\toprule
Theorem & Lines & Deps \\
\midrule
\endhead

\bottomrule
\endfoot

\mgw{Tietze_stepII_real_extension_nonempty} & 10369 & 142 \\
\mgw{Urysohn_lemma} & 2964 & 85 \\
\mgw{Urysohn_metrization_theorem} & 2174 & 112 \\
\mgw{one_point_compactification_exists} & 1719 & 38 \\
\mgw{closed_interval_affine_equiv_minus1_1} & 1638 & 48 \\
\mgw{order_topology_basis_is_basis} & 1164 & 7 \\
\mgw{order_rel_trans} & 1159 & 22 \\
\mgw{unit_interval_connected} & 1049 & 32 \\
\mgw{uniform_topology_finer_than_product_and_coarser_than_box} & 940 & 74 \\
\mgw{convex_subspace_order_topology} & 931 & 26 \\
\mgw{ordered_square_not_subspace_dictionary} & 922 & 42 \\
\mgw{countability_axioms_subspace_product} & 911 & 51 \\
\mgw{product_topology_full_regular_axiom} & 907 & 56 \\
\mgw{rectangular_regions_basis_plane} & 894 & 12 \\
\mgw{interval_connected} & 878 & 23 \\
\mgw{completely_regular_product_topology} & 876 & 56 \\
\mgw{bounded_transform_psi_preimage_open_ray_upper} & 812 & 26 \\
\mgw{Romega_D_metric_open_ball_in_product_topology} & 805 & 62 \\
\mgw{ball_inside_rectangle} & 797 & 20 \\
\mgw{ex13_8b_halfopen_rational_basis_topology} & 768 & 18 \\
\mgw{Sorgenfrey_line_Lindelof} & 756 & 52 \\
\mgw{ex16_9_dictionary_equals_product} & 736 & 55 \\
\mgw{pointwise_limit_continuity_points_dense} & 729 & 58 \\
\mgw{ex13_8a_rational_intervals_basis_standard} & 710 & 16 \\
\mgw{ex16_10_compare_topologies_on_square} & 693 & 40 \\
\mgw{continuous_combine_or_01} & 677 & 43 \\
\mgw{metrizable_spaces_normal} & 674 & 37 \\
\mgw{two_by_nat_singleton_not_in_basis} & 664 & 8 \\
\mgw{double_minus_one_map_continuous} & 629 & 43 \\
\mgw{embedding_via_functions} & 625 & 52 \\
\mgw{compact_space_net_has_accumulation_point} & 615 & 23 \\
\mgw{regular_countable_basis_normal} & 600 & 28 \\
\mgw{compact_space_net_has_accumulation_point_on} & 590 & 23 \\
\mgw{R_standard_basis_is_basis_local} & 585 & 9 \\
\mgw{euclidean_spaces_second_countable} & 581 & 48 \\
\mgw{preimage_mul_fun_R_open_ray_upper_in_product_topology} & 579 & 42 \\
\mgw{Tietze_extension_open_interval} & 576 & 48 \\
\mgw{Romega_extend_map_continuous_in_Romega_product} & 574 & 47 \\
\mgw{product_countable_basis_at_point_if_components_first_countable} & 574 & 34 \\
\mgw{abs_Cauchy_sequence_converges_R_standard_topology} & 563 & 30 \\
\mgw{order_rel_trichotomy_or_impred} & 560 & 13 \\
\mgw{preimage_mul_fun_R_open_ray_lower_in_product_topology} & 551 & 41 \\
\mgw{R_upper_limit_basis_is_basis_local} & 534 & 10 \\
\mgw{Q_sqrt2_cut_not_interval_or_ray} & 533 & 31 \\
\mgw{abs_Cauchy_sequence_converges_R_standard_topology_early} & 531 & 30 \\
\mgw{Sorgenfrey_plane_not_normal} & 525 & 33 \\
\mgw{singleton_rectangle_in_dictionary} & 514 & 45 \\
\mgw{subnet_converges_to_accumulation} & 495 & 32 \\
\mgw{subnet_converges_to_accumulation_witnessed} & 487 & 32 \\
\mgw{open_ball_refine_intersection} & 481 & 17 \\
\mgw{continuous_construction_rules} & 475 & 16 \\
\mgw{ex17_5_basis_elem_meets_interval} & 474 & 10 \\
\mgw{flip_unit_interval_continuous} & 473 & 36 \\
\mgw{uniform_cauchy_metric_complete_imp_uniform_limit_stub} & 467 & 30 \\
\mgw{subnet_converges_to_accumulation_witnessed_on} & 465 & 32 \\
\mgw{A_N_eps_closed_stub} & 463 & 32 \\
\mgw{paracompact_Hausdorff_normal} & 460 & 28 \\
\mgw{Tietze_extension_minus1_1} & 457 & 48 \\
\mgw{product_sequence_convergence_iff_coordinates} & 456 & 42 \\
\mgw{R_lower_limit_basis_is_basis_local} & 454 & 9 \\
\mgw{unbounded_sequences_in_Romega_box_topology} & 437 & 27 \\
\mgw{preimage_add_fun_R_open_ray_upper_in_product_topology} & 435 & 32 \\
\mgw{ex30_4_compact_metrizable_second_countable} & 425 & 48 \\
\mgw{regular_normal_via_closure} & 421 & 14 \\
\mgw{limit_point_compact_not_necessarily_compact} & 421 & 25 \\
\mgw{separation_subspace_limit_points} & 419 & 9 \\
\mgw{paracompact_Hausdorff_regular} & 416 & 25 \\
\mgw{one_minus_fun_continuous} & 413 & 32 \\
\mgw{ex13_6_Rl_RK_not_comparable} & 411 & 24 \\
\mgw{double_map_continuous} & 411 & 41 \\
\mgw{distance_R2_triangle_sqr_Le} & 409 & 14 \\
\mgw{Cauchy_with_convergent_subsequence_converges} & 404 & 41 \\
\mgw{finite_product_compact} & 402 & 21 \\
\mgw{Romega_product_cylinder_in_D_metric_topology} & 399 & 37 \\
\mgw{R_standard_plus_K_basis_is_basis_local} & 397 & 9 \\
\mgw{continuity_via_nets} & 383 & 33 \\
\mgw{Sorgenfrey_plane_L_closed} & 381 & 34 \\
\mgw{Heine_Borel_closed_bounded} & 380 & 26 \\
\mgw{closed_subspace_paracompact} & 378 & 16 \\
\mgw{R_lub_exists} & 376 & 13 \\
\mgw{continuity_via_sequences_metric} & 374 & 21 \\
\mgw{ex17_17_closure_A_C_Supq} & 370 & 19 \\
\mgw{Romega_singleton_map_continuous_prod} & 367 & 31 \\
\mgw{ex30_14_product_Lindelof_compact} & 362 & 28 \\
\mgw{product_finite_subbasis_intersection_separator} & 357 & 43 \\
\mgw{uniform_limit_of_continuous_to_metric_is_continuous} & 352 & 35 \\
\mgw{ex30_5a_metrizable_countable_dense_second_countable} & 347 & 39 \\
\mgw{ex23_connected_open_sets_path_connected} & 347 & 27 \\
\mgw{completely_regular_subspace_product} & 346 & 49 \\
\mgw{rectangle_inside_ball} & 345 & 25 \\
\mgw{preimage_add_fun_R_open_ray_lower_in_product_topology} & 341 & 32 \\
\mgw{path_component_transitive_axiom} & 335 & 41 \\
\mgw{compact_subspace_via_ambient_covers} & 335 & 8 \\
\mgw{locally_m_euclidean_implies_T1} & 330 & 25 \\
\mgw{ex30_5b_metrizable_Lindelof_second_countable} & 328 & 34 \\
\mgw{Romega_D_metric_induces_product_topology} & 327 & 46 \\
\mgw{product_basis_generates} & 322 & 10 \\
\mgw{U_eps_open_dense_stub} & 320 & 38 \\
\mgw{RK_not_regular_axiom} & 320 & 23 \\
\mgw{Sigma_countable} & 318 & 2 \\
\mgw{order_topology_interval_refine} & 313 & 5 \\
\mgw{topology_chain_four_sets} & 310 & 4 \\
\mgw{pasting_lemma} & 306 & 18 \\
\mgw{ex13_7_R_topology_containments} & 304 & 23 \\
\mgw{Sorgenfrey_plane_diag_rational_closed} & 303 & 33 \\
\mgw{finite_product_connected} & 303 & 17 \\
\mgw{compact_Hausdorff_normal} & 301 & 17 \\

\end{longtable}

Table~\ref{tab:chatgpt30} shows an assessment of 30 major theorems/lemmas selected by ChatGPT. It may be incorrect in various ways. The exact statements of these (and some more) theorems/lemmas in LaTeX and Megalodon are shown in the Appendix.
\begin{longtable}{|p{2.6cm}|p{2.8cm}|p{3.1cm}|p{4.9cm}|}
\caption{ChatGPT selection and assessment of 30 major results}\label{tab:chatgpt30} \\
\hline
\textbf{Section \& No. (topology.tex)} &
\textbf{Theorem / Lemma name} &
\textbf{Formal item} &
\textbf{ChatGPT Assessment of formalization} \\
\hline
\endfirsthead

\hline
\textbf{Section \& No.} &
\textbf{Theorem / Lemma} &
\textbf{Formal item (:T)} &
\textbf{Assessment} \\
\hline
\endhead

\hline
\endfoot

\hline
\endlastfoot

§15, Thm.\ 15.1 &
Basis for product topology &
\formal{product\_\allowbreak basis\_\allowbreak generates\_\allowbreak product\_\allowbreak topology} &
Clean abstraction of basis arguments; faithful to textbook proof, good reuse of basis lemmas. \\
\hline

§16, Thm.\ 16.3 &
Subspace of a product &
\formal{product\_\allowbreak subspace\_\allowbreak topology} &
Correct handling of induced topology; explicit but robust. \\
\hline

§18, Thm.\ 18.3 &
Pasting Lemma &
\formal{pasting\_\allowbreak lemma} &
Excellent stress test for continuity infrastructure; hypotheses well-scoped. \\
\hline

§18, Thm.\ 18.4 &
Maps into products &
\formal{maps\_\allowbreak into\_\allowbreak products} &
Very standard statement, formalization mirrors textbook closely. \\
\hline

§23, Lem.\ 23.1 &
Connectedness via separations &
\formal{connected\_\allowbreak space\_\allowbreak no\_\allowbreak separation} &
Correct equivalence formulation; slightly low-level set reasoning but sound. \\
\hline

§23, Lem.\ 23.2 &
Connected subset and separations &
\formal{connected\_\allowbreak subset\_\allowbreak in\_\allowbreak separation\_\allowbreak side} &
Faithful encoding; reusable lemma for later proofs. \\
\hline

§23, Thm.\ 23.3 &
Union of connected sets &
\formal{union\_\allowbreak connected\_\allowbreak common\_\allowbreak point} &
Good abstraction, proof structure matches standard argument. \\
\hline

§23, Thm.\ 23.5 &
Continuous image of connected space &
\formal{continuous\_\allowbreak image\_\allowbreak connected} &
Canonical theorem; formal statement is clean and general. \\
\hline

§23, Thm.\ 23.6 &
Finite products of connected spaces &
\formal{finite\_\allowbreak product\_\allowbreak connected} &
Correct induction on finite products; abstraction level appropriate. \\
\hline

§26, Thm.\ 26.2 &
Closed subspace of compact space &
\formal{closed\_\allowbreak subspace\_\allowbreak compact} &
Very standard; good foundation for later compactness results. \\
\hline

§26, Thm.\ 26.3 &
Compact subsets of Hausdorff spaces &
\formal{compact\_\allowbreak subspace\_\allowbreak in\_\allowbreak Hausdorff\_\allowbreak closed} &
Faithful separation argument; predicates well chosen. \\
\hline

§26, Thm.\ 26.5 &
Continuous image of compact space &
\formal{continuous\_\allowbreak image\_\allowbreak compact} &
Essential lemma; well-factored and reusable. \\
\hline

§26, Thm.\ 26.6 &
Compact--Hausdorff bijection &
\formal{compact\_\allowbreak to\_\allowbreak Hausdorff\_\allowbreak bijection\_\allowbreak homeomorphism} &
Excellent formulation; topologically canonical. \\
\hline

§26, Thm.\ 26.7 &
Finite products of compact spaces &
\formal{finite\_\allowbreak product\_\allowbreak compact} &
Nontrivial induction; proof infrastructure appears solid. \\
\hline

§26, Lem.\ 26.8 &
Tube Lemma &
\formal{tube\_\allowbreak lemma} &
Good handling of product neighborhoods; faithful to Munkres. \\
\hline

§27, Thm.\ 27.3 &
Heine--Borel Theorem &
\formal{Heine\_\allowbreak Borel\_\allowbreak closed\_\allowbreak bounded} &
Major result; hypotheses and equivalence encoded correctly. \\
\hline

§27, Lem.\ 27.5 &
Lebesgue Number Lemma &
\formal{Lebesgue\_\allowbreak number\_\allowbreak lemma} &
Good cover machinery; slightly verbose but robust. \\
\hline

§27, Thm.\ 27.6 &
Uniform continuity on compact spaces &
\formal{uniform\_\allowbreak continuity\_\allowbreak on\_\allowbreak compact} &
Correct metric-topology interaction; good abstraction. \\
\hline

§30, Thm.\ 30.3 &
Countable basis $\Rightarrow$ Lindelöf &
\formal{countable\_\allowbreak basis\_\allowbreak implies\_\allowbreak Lindelof} &
Nicely structured countability argument. \\
\hline

§30, Thm.\ 30.4 &
Countable basis $\Rightarrow$ separable &
\formal{countable\_\allowbreak basis\_\allowbreak implies\_\allowbreak separable} &
Faithful, concise; good reuse of basis machinery. \\
\hline

§32, Thm.\ 32.1 &
Regular + countable basis $\Rightarrow$ normal &
\formal{regular\_\allowbreak countable\_\allowbreak basis\_\allowbreak normal} &
Important separation result; dependencies are appropriate. \\
\hline

§32, Thm.\ 32.3 &
Compact Hausdorff spaces are normal &
\formal{compact\_\allowbreak Hausdorff\_\allowbreak normal} &
Standard argument; formalization is clean. \\
\hline

§33, Thm.\ 33.1 &
Urysohn Lemma &
\formal{Urysohn\_\allowbreak lemma} &
Excellent: correct hypotheses, nontrivial construction handled well. \\
\hline

§34, Thm.\ 34.1 &
Urysohn Metrization Theorem &
\formal{Urysohn\_\allowbreak metrization\_\allowbreak theorem} &
High-value result; formalization appears faithful and well-structured. \\
\hline

§35, Thm.\ 35.1 &
Tietze Extension Theorem &
\formal{Tietze\_\allowbreak extension\_\allowbreak interval} &
Very strong result; formal statement and dependencies look sound. \\
\hline

§41, Thm.\ 41.1 &
Paracompact Hausdorff $\Rightarrow$ normal &
\formal{paracompact\_\allowbreak Hausdorff\_\allowbreak normal} &
Good abstraction of locally finite covers. \\
\hline

§41, Thm.\ 41.2 &
Closed subspaces of paracompact spaces &
\formal{closed\_\allowbreak subspace\_\allowbreak paracompact} &
Correct and reusable; well integrated. \\
\hline

§41, Lem.\ 41.3 &
Michael Lemma &
\formal{Michael\_\allowbreak lemma\_\allowbreak 41\_\allowbreak 3} &
Technically demanding; solid formal handling. \\
\hline

§41, Lem.\ 41.6 &
Shrinking Lemma &
\formal{shrinking\_\allowbreak lemma\_\allowbreak 41\_\allowbreak 6} &
Good control of refinements; abstraction is adequate. \\
\hline

§48, Lem.\ 48.1 &
Baire space lemma &
\formal{Baire\_\allowbreak space\_\allowbreak dense\_\allowbreak Gdelta} &
Correct use of countable intersections; solid foundation for category arguments. \\

\hline

\end{longtable}

\section{The Battle of the Tietze Hill}

A large part of the last two days has been spent on the formalization
of the Tietze extension theorem, specifically its Step II. Its (autoformalized) Megalodon statement is as follows:
\begin{small}
\begin{verbatim}
(** helper for §35 Step II: nonempty closed subset case, real-valued extension **)
(** LATEX VERSION: Step II (nonempty A): construct a real-valued continuous
    extension gR:X->R agreeing with f on A and bounded in [-1,1]. **)
Theorem Tietze_stepII_real_extension_nonempty : forall X Tx A f:set,
  normal_space X Tx ->
  closed_in X Tx A ->
  A <> Empty ->
  continuous_map A (subspace_topology X Tx A)
    (closed_interval (minus_SNo 1) 1) (closed_interval_topology (minus_SNo 1) 1) f ->
  exists gR:set,
    continuous_map X Tx R R_standard_topology gR /\
    (forall x:set, x :e A -> apply_fun gR x = apply_fun f x) /\
    (forall x:set, x :e X -> apply_fun gR x :e closed_interval (minus_SNo 1) 1).
\end{verbatim}
\end{small}  
The proof was finished on January 4 at 20:15, with the following (pretty dry given the effort) progress message about backup nr 3893:
\begin{small}
\begin{verbatim}
Update 3893: §35 Step II: eliminated the temporary admit in Htail_budget by adding
the missing `HgN0xS` (for the current x0) and closing the final SNoLe_tra transitivity;
Tietze_stepII_real_extension_nonempty now has no admits.
\end{verbatim}
\end{small}
The proof\footnote{\url{https://mgwiki.github.io/mgw_test/topology.mg.html\#Tietze_stepII_real_extension_nonempty} , also line 129871 of \url{https://raw.githubusercontent.com/mgwiki/mgw_test/48bca1fc5df4576f1a28416aa4e423ecfa556f6c/mglib/topology.mg}} takes 10369 lines and has 142 direct dependencies (36 definitions and 106 theorems).

The textbook proof uses
facts about infinite sums of real functions and their uniform
convergence, which is a major external dependency not present in the
textbook. The agent seems to have been stumped by this and instead
kept unrolling the inductive (infinite) construction 12 times (12
terms of the series), producing about 6k lines that are not useful for
the final proof (even though they may be useful for figuring out the
machinery more concretely before the induction is done).

It has however recovered from this and decided to gradually formalize the necessary prerequisites.
The ChatGPT analysis of the unfinished 9k-line proof attempt is available online\footnote{\url{https://chatgpt.com/share/695a7928-9efc-8002-860f-bb2c0f33663b}}. Below is its periodic report, showing how the fact that complete metric spaces are dealt with much later in the textbook confused the LLM.
\begin{small}
\begin{verbatim}
CHANGES3682:
Attempted refactor to move the uniform-limit/uniform-Cauchy
infrastructure (continuous_at_map, pointwise_limit_metric,
uniform_limit_metric, uniform_cauchy_metric, and the uniform-limit
continuity lemmas) earlier near §35 to support Tietze Step II.

Result:
- This does not work because that block depends on
`complete_metric_space`, which is defined later (in the metric
completeness section), so moving it before §35 causes a
forward-reference error: "Unknown term complete_metric_space".

Action taken:
- Reverted Math_Background.mg back to bck3681 state (no content loss;
only the failed move was undone).
\end{verbatim}
\end{small}
In more detail, the problem is that complete metric spaces and Cauchy sequences in them are indeed only defined in section 43 of Munkres, while Tietze is in section 35. The LLM has tried hard to follow the formalization rules. 
As a result, it has done 11 useless iterations and correction terms (from u2 in CHANGES3650 up to u12 in CHANGES3685), blowing up the proof from 865 lines to 5408 lines in these 35 changes. This took almost 4 hours of work and wasted quite some resources.
After some futile attempts, the LLM has reasoned itself into doing the hard work on the prerequisites and moving them before Tietze as follows:
\begin{small}
\begin{verbatim}
CHANGES3687:
Move uniform-limit infrastructure before §35 (compiles)
======================================================

Goal:
- §35 (Tietze extension theorem, Step II) needs uniform-limit /
completeness tools without forward-reference blockers.

What changed:
- Moved the block starting at "helper: continuity at a point into a
topological space" (continuous_at_map, pointwise_limit_metric,
uniform_limit_metric, uniform_cauchy_metric, uniform-limit continuity
lemmas, plus related helpers) from its later location to the area just
before §35.
- Kept §48-specific helper material (A_N_eps, U_eps,
U_eps_open_dense_stub, etc.) located near §48 again, so it stays after
the Baire_space definitions.
- Added early stubs needed by the moved uniform-limit block:
  - metric_topology_R_bounded_metric_eq_R_standard_topology_early
  - R_bounded_metric_is_metric_on_early
  These avoid forward references to the later bounded-metric
  development; they should later be replaced by the proved theorems.
- Left a note in the original location indicating the helper block was moved earlier.
\end{verbatim}
\end{small}
Note that this qualifies as a large gap, external dependency or even an (unsafe) forward reference in the
textbook -- something that can be a serious hurdle for human formalizers.
The ChatGPT analysis of the final 10k-line proof and its difference to the previous unfinished 9k-line proof is available online\footnote{\url{https://chatgpt.com/share/695a7928-9efc-8002-860f-bb2c0f33663b}}. 
Once this hard battle is finished, the standard statement of Tietze from Munkres which is as follows:
\begin{center}
\fbox{%
\parbox{\textwidth}{%
\textbf{Theorem 35.1 (Tietze extension theorem).}
Let $X$ be a normal space; let $A$ be a closed subspace of $X$.

\medskip
(a) Any continuous map of $A$ into the closed interval $[a,b]$ of $\mathbb{R}$
may be extended to a continuous map of all of $X$ into $[a,b]$.

\medskip
(b) Any continuous map of $A$ into $\mathbb{R}$
may be extended to a continuous map of all of $X$ into $\mathbb{R}$.
}%
}
\end{center}
is (relatively easily) proved via these two Megalodon theorems:
\begin{small}
\begin{verbatim}
Theorem Tietze_extension_real_bounded_interval : forall X Tx A a b f:set,
  Rle a b ->
  normal_space X Tx -> closed_in X Tx A ->
  continuous_map A (subspace_topology X Tx A) R R_standard_topology f ->
  (forall x:set, x :e A -> apply_fun f x :e closed_interval a b) ->
  exists g:set, continuous_map X Tx R R_standard_topology g /\
    (forall x:set, x :e A -> apply_fun g x = apply_fun f x).

Theorem Tietze_extension_real : forall X Tx A f:set,
  normal_space X Tx -> closed_in X Tx A ->
  continuous_map A (subspace_topology X Tx A) R R_standard_topology f ->
  exists g:set, continuous_map X Tx R R_standard_topology g /\
    (forall x:set, x :e A -> apply_fun g x = apply_fun f x).
\end{verbatim}
\end{small}

\section{Conclusion}
This is obviously quite surprising and exciting and I wonder where this will all
go. The price tag here (and my coding effort) was really low. It is quite possible that in 2026 we will get most of (reasonably written)
math textbooks and papers autoformalized. Or perhaps not - there may be limits
this will eventually run into. Perhaps a (sorely needed?) reality check for some
(insert a popular proof assistant name here) fundamentalists is that the current
experiments were done in an almost unknown proof assistant based on higher-order
set theory, with very little exposure to the existing LLMs. The future of
AI-assisted (auto)formalization seems to be quite open and democratic.

\section{Acknowledgments}
Chad Brown has largely developed Megalodon, its hammer, and the
initial core set-theoretical library, helped me to set it up for the
experiment, and provided encouraging comments on the early system's
formalizations. Zar Goertzel has kept me informed about his
experiments with LLMs and the related coding agents like Claude Code
and Codex. I took his advice and fully switched to ChatGPT 5.2 (and
the \$200/month Pro subscription) after he reported encouraging
results. He has already finished a 13k-line autoformalization of the
Ramsey(3,6) theorem. Atle Hahn and Adrian De Lon have been doing
interesting experiments with LLM-based autoformalization into
Naproche. Atle's impressive (and working) prompts demonstrate the
power and potential of "programming" LLMs to do
autoformalization.
Thanks to John Harrison for supporting our early autoformalization efforts and also to Larry Paulson and Dominic Mulligan for encouragement.
Cezary Kaliszyk has helped with the
Megalodon/hammer development and related infrastructure, and
spearheaded the first learning-based autoformalization ideas, projects
and approaches with me since 2014. 
The autoformalization community has
grown a lot since then - thanks to all these (and especially the
early) believers.

\bibliography{afgpt}

\begin{thebibliography}{10}

\bibitem{DBLP:conf/mkm/BrownKSU25}
Chad~E. Brown, Cezary Kaliszyk, Martin Suda, and Josef Urban.
\newblock Hammering higher order set theory.
\newblock In Valeria de~Paiva and Peter Koepke, editors, {\em Intelligent
  Computer Mathematics - 18th International Conference, {CICM} 2025, Brasilia,
  Brazil, October 6-10, 2025, Proceedings}, volume 16136 of {\em Lecture Notes
  in Computer Science}, pages 3--20. Springer, 2025.
\newblock \href {https://doi.org/10.1007/978-3-032-07021-0\_1}
  {\path{doi:10.1007/978-3-032-07021-0\_1}}.

\bibitem{cariolaro1999ramsey}
David Cariolaro.
\newblock {\em On the Ramsey number R (3, 6)}.
\newblock Department of Mathematical Sciences, Aalborg University, 1999.

\bibitem{DBLP:journals/ijar/GauthierOU23}
Thibault Gauthier, Miroslav Ols{\'{a}}k, and Josef Urban.
\newblock Alien coding.
\newblock {\em Int. J. Approx. Reason.}, 162:109009, 2023.
\newblock URL: \url{https://doi.org/10.1016/j.ijar.2023.109009}, \href
  {https://doi.org/10.1016/J.IJAR.2023.109009}
  {\path{doi:10.1016/J.IJAR.2023.109009}}.

\bibitem{DBLP:conf/cade/GauthierU25}
Thibault Gauthier and Josef Urban.
\newblock Learning conjecturing from scratch.
\newblock In Clark~W. Barrett and Uwe Waldmann, editors, {\em Automated
  Deduction - {CADE} 30 - 30th International Conference on Automated Deduction,
  Stuttgart, Germany, July 28-31, 2025, Proceedings}, volume 15943 of {\em
  Lecture Notes in Computer Science}, pages 423--445. Springer, 2025.
\newblock \href {https://doi.org/10.1007/978-3-031-99984-0\_23}
  {\path{doi:10.1007/978-3-031-99984-0\_23}}.

\bibitem{DBLP:conf/itp/JakubuvCGKOP00U23}
Jan Jakubuv, Karel Chvalovsk{\'{y}}, Zarathustra~Amadeus Goertzel, Cezary
  Kaliszyk, Mirek Ols{\'{a}}k, Bartosz Piotrowski, Stephan Schulz, Martin Suda,
  and Josef Urban.
\newblock Mizar 60 for mizar 50.
\newblock In Adam Naumowicz and Ren{\'{e}} Thiemann, editors, {\em 14th
  International Conference on Interactive Theorem Proving, {ITP} 2023,
  Bia{\l}ystok, Poland, July 31 - August 4, 2023}, volume 268 of {\em LIPIcs},
  pages 19:1--19:22. Schloss Dagstuhl - Leibniz-Zentrum f{\"{u}}r Informatik,
  2023.
\newblock URL: \url{https://doi.org/10.4230/LIPIcs.ITP.2023.19}, \href
  {https://doi.org/10.4230/LIPICS.ITP.2023.19}
  {\path{doi:10.4230/LIPICS.ITP.2023.19}}.

\bibitem{DBLP:conf/itp/KaliszykUV15}
Cezary Kaliszyk, Josef Urban, and Jir{\'{\i}} Vyskocil.
\newblock Learning to parse on aligned corpora (rough diamond).
\newblock In Christian Urban and Xingyuan Zhang, editors, {\em Interactive
  Theorem Proving - 6th International Conference, {ITP} 2015, Nanjing, China,
  August 24-27, 2015, Proceedings}, volume 9236 of {\em Lecture Notes in
  Computer Science}, pages 227--233. Springer, 2015.
\newblock \href {https://doi.org/10.1007/978-3-319-22102-1\_15}
  {\path{doi:10.1007/978-3-319-22102-1\_15}}.

\bibitem{DBLP:conf/mkm/KaliszykUVG14}
Cezary Kaliszyk, Josef Urban, Jir{\'{\i}} Vyskocil, and Herman Geuvers.
\newblock Developing corpus-based translation methods between informal and
  formal mathematics: Project description.
\newblock In Stephen~M. Watt, James~H. Davenport, Alan~P. Sexton, Petr Sojka,
  and Josef Urban, editors, {\em Intelligent Computer Mathematics -
  International Conference, {CICM} 2014, Coimbra, Portugal, July 7-11, 2014.
  Proceedings}, volume 8543 of {\em Lecture Notes in Computer Science}, pages
  435--439. Springer, 2014.
\newblock \href {https://doi.org/10.1007/978-3-319-08434-3\_34}
  {\path{doi:10.1007/978-3-319-08434-3\_34}}.

\bibitem{DBLP:journals/jsc/PiepenbrockUKOHJ25}
Jelle Piepenbrock, Josef Urban, Konstantin Korovin, Miroslav Ols{\'{a}}k, Tom
  Heskes, and Mikol{\'{a}}s Janota.
\newblock Invariant neural architecture for learning term synthesis in
  instantiation proving.
\newblock {\em J. Symb. Comput.}, 128:102375, 2025.
\newblock URL: \url{https://doi.org/10.1016/j.jsc.2024.102375}, \href
  {https://doi.org/10.1016/J.JSC.2024.102375}
  {\path{doi:10.1016/J.JSC.2024.102375}}.

\bibitem{DBLP:conf/cade/Urban07}
Josef Urban.
\newblock Malarea: a metasystem for automated reasoning in large theories.
\newblock In Geoff Sutcliffe, Josef Urban, and Stephan Schulz, editors, {\em
  Proceedings of the {CADE-21} Workshop on Empirically Successful Automated
  Reasoning in Large Theories, Bremen, Germany, 17th July 2007}, volume 257 of
  {\em {CEUR} Workshop Proceedings}. CEUR-WS.org, 2007.
\newblock URL: \url{https://ceur-ws.org/Vol-257/05\_Urban.pdf}.

\bibitem{DBLP:conf/lpar/Urban08}
Josef Urban.
\newblock Automated reasoning for mizar: Artificial intelligence through
  knowledge exchange.
\newblock In Piotr Rudnicki, Geoff Sutcliffe, Boris Konev, Renate~A. Schmidt,
  and Stephan Schulz, editors, {\em Proceedings of the {LPAR} 2008 Workshops,
  Knowledge Exchange: Automated Provers and Proof Assistants, and the 7th
  International Workshop on the Implementation of Logics, Doha, Qatar, November
  22, 2008}, volume 418 of {\em {CEUR} Workshop Proceedings}. CEUR-WS.org,
  2008.
\newblock URL: \url{https://ceur-ws.org/Vol-418/paper1.pdf}.

\bibitem{DBLP:conf/mkm/UrbanJ20}
Josef Urban and Jan Jakubuv.
\newblock First neural conjecturing datasets and experiments.
\newblock In Christoph Benzm{\"{u}}ller and Bruce~R. Miller, editors, {\em
  Intelligent Computer Mathematics - 13th International Conference, {CICM}
  2020, Bertinoro, Italy, July 26-31, 2020, Proceedings}, volume 12236 of {\em
  Lecture Notes in Computer Science}, pages 315--323. Springer, 2020.
\newblock \href {https://doi.org/10.1007/978-3-030-53518-6\_24}
  {\path{doi:10.1007/978-3-030-53518-6\_24}}.

\bibitem{DBLP:conf/mkm/WangKU18}
Qingxiang Wang, Cezary Kaliszyk, and Josef Urban.
\newblock First experiments with neural translation of informal to formal
  mathematics.
\newblock In Florian Rabe, William~M. Farmer, Grant~O. Passmore, and Abdou
  Youssef, editors, {\em Intelligent Computer Mathematics - 11th International
  Conference, {CICM} 2018, Hagenberg, Austria, August 13-17, 2018,
  Proceedings}, volume 11006 of {\em Lecture Notes in Computer Science}, pages
  255--270. Springer, 2018.
\newblock \href {https://doi.org/10.1007/978-3-319-96812-4\_22}
  {\path{doi:10.1007/978-3-319-96812-4\_22}}.

\end{thebibliography}

\appendix

\section{Listing of the Major Theorems in LaTeX and Megalodon}

\begin{longtable}{|p{1.0cm}|p{4.2cm}|p{2.5cm}|p{5.8cm}|}
\hline
\textbf{No.} & \textbf{Statement in \texttt{topology.tex}} & \textbf{Formal name} & \textbf{Formal statement (stats, exact)} \\
\hline
\endfirsthead
\hline
\textbf{No.} & \textbf{Statement in \texttt{topology.tex}} & \textbf{Formal name} & \textbf{Formal statement (stats, exact)} \\
\hline
\endhead
\hline
\endfoot
\hline
\endlastfoot

Th~15.1 &
Th 15.1. If $B$ is a basis for the topology of $X$ and $\mathcal{C}$ is a basis for the topology of $Y$, then the collection $\mathscr{B} \times \mathcal{C}=\{B \times C \mid B \in \mathscr{B} \text { and } C \in \mathcal{C}\}$ is a basis for the topology of $X \times Y$. &
\texttt{product\_\allowbreak basis\_\allowbreak generates\_product\_\allowbreak topology} &
\Verb[breaklines=true,breakanywhere=true]!forall X Y Bx By:set, basis_on X Bx -> basis_on Y By -> product_topology X (generated_topology X Bx) Y (generated_topology Y By) = generated_topology (setprod X Y) (product_basis_from X Y Bx By)! \\
\hline

Th~16.3 &
Th 16.3. If $A$ is a subspace of $X$ and $B$ is a subspace of $Y$, then the product topology on $A \times B$ is the same as the topology $A \times B$ inherits as a subspace of $X \times Y$. &
\texttt{product\_\allowbreak subspace\_\allowbreak topology} &
\Verb[breaklines=true,breakanywhere=true]!forall X Tx Y Ty A B:set, topology_on X Tx -> topology_on Y Ty -> product_topology A (subspace_topology X Tx A) B (subspace_topology Y Ty B) = subspace_topology (setprod X Y) (product_topology X Tx Y Ty) (setprod A B)! \\
\hline

Th~18.3 &
Th 18.3 (Pasting lemma). Suppose that $X=A \cup B$, where $A$ and $B$ are closed in $X$. Let $f: X \rightarrow Y$ be a map such that $f \mid A$ and $f \mid B$ are continuous. If $f \mid A \cap B$ is continuous, then $f$ is continuous. &
\texttt{pasting\_\allowbreak lemma} &
\Verb[breaklines=true,breakanywhere=true]!forall X Tx Y Ty A B f:set, topology_on X Tx -> topology_on Y Ty -> closed_in X Tx A -> closed_in X Tx B -> A :\/: B = X -> continuous_map A (subspace_topology X Tx A) Y Ty (restriction_map A f) -> continuous_map B (subspace_topology X Tx B) Y Ty (restriction_map B f) -> (forall x:set, x :e (A :/\: B) -> apply_fun f x = apply_fun f x) -> continuous_map X Tx Y Ty f! \\
\hline

Th~18.4 &
Th 18.4. Let $f: X \rightarrow Y \times Z$ be given by the equation $f(x)=(f_{1}(x), f_{2}(x))$. Then $f$ is continuous if and only if $f_{1}$ and $f_{2}$ are continuous. &
\texttt{maps\_\allowbreak into\_\allowbreak products} &
\Verb[breaklines=true,breakanywhere=true]!forall A Ta X Tx Y Ty h:set, topology_on A Ta -> topology_on X Tx -> topology_on Y Ty -> (continuous_map A Ta (setprod X Y) (product_topology X Tx Y Ty) h <-> (continuous_map A Ta X Tx (projection_map1 X Y o h) /\ continuous_map A Ta Y Ty (projection_map2 X Y o h)))! \\
\hline

Lem~23.2 &
Lem 23.2. If $C$ is a connected subspace of $X$ that intersects both sets $A$ and $B$ of a separation $X=A \cup B$, then $C$ intersects $A \cap B$. &
\texttt{connected\_\allowbreak subset\_\allowbreak in\_separation\_\allowbreak side} &
\Verb[breaklines=true,breakanywhere=true]!forall X Tx C D A:set, topology_on X Tx -> connected_space C (subspace_topology X Tx C) -> separation_of X Tx A D -> C :/\: A <> Empty -> C :/\: D <> Empty -> C :/\: A :/\: D <> Empty! \\
\hline

Th~23.3 &
Th 23.3. The union of a collection of connected subspaces of $X$ that have a point in common is connected. &
\texttt{union\_\allowbreak connected\_\allowbreak common\_\allowbreak point} &
\Verb[breaklines=true,breakanywhere=true]!forall X Tx F x0:set, topology_on X Tx -> (forall C:set, C :e F -> connected_space C (subspace_topology X Tx C)) -> (forall C:set, C :e F -> x0 :e C) -> connected_space (Union F) (subspace_topology X Tx (Union F))! \\
\hline

Th~23.5 &
Th 23.5. The image of a connected space under a continuous map is connected. &
\texttt{continuous\_\allowbreak image\_\allowbreak connected} &
\Verb[breaklines=true,breakanywhere=true]!forall X Tx Y Ty f:set, topology_on X Tx -> topology_on Y Ty -> continuous_map X Tx Y Ty f -> connected_space X Tx -> connected_space (image_of X f) (subspace_topology Y Ty (image_of X f))! \\
\hline

Th~23.6 &
Th 23.6. The product of finitely many connected spaces is connected. &
\texttt{finite\_\allowbreak product\_\allowbreak connected} &
\Verb[breaklines=true,breakanywhere=true]!forall X Tx Y Ty:set, connected_space X Tx -> connected_space Y Ty -> connected_space (setprod X Y) (product_topology X Tx Y Ty)! \\
\hline

Th~26.2 &
Th 26.2. Every closed subspace of a compact space is compact. &
\texttt{closed\_\allowbreak subspace\_\allowbreak compact} &
\Verb[breaklines=true,breakanywhere=true]!forall X Tx Y:set, compact_space X Tx -> closed_in X Tx Y -> compact_space Y (subspace_topology X Tx Y)! \\
\hline

Th~26.3 &
Th 26.3. Every compact subspace of a Hausdorff space is closed. &
\texttt{compact\_\allowbreak subspace\_\allowbreak in\_Hausdorff\_\allowbreak closed} &
\Verb[breaklines=true,breakanywhere=true]!forall X Tx Y:set, Hausdorff_space X Tx -> compact_space Y (subspace_topology X Tx Y) -> closed_in X Tx Y! \\
\hline

Th~26.5 &
Th 26.5. The image of a compact space under a continuous map is compact. &
\texttt{continuous\_\allowbreak image\_\allowbreak compact} &
\Verb[breaklines=true,breakanywhere=true]!forall X Tx Y Ty f:set, compact_space X Tx -> topology_on Y Ty -> continuous_map X Tx Y Ty f -> compact_space (image_of X f) (subspace_topology Y Ty (image_of X f))! \\
\hline

Th~26.6 &
Th 26.6. Let $f: X \rightarrow Y$ be a bijective continuous function. If $X$ is compact and $Y$ is Hausdorff, then $f$ is a homeomorphism. &
\texttt{compact\_\allowbreak to\_\allowbreak Hausdorff\_bijection\_\allowbreak homeomorphism} &
\Verb[breaklines=true,breakanywhere=true]!forall X Tx Y Ty f:set, compact_space X Tx -> Hausdorff_space Y Ty -> bijection X Y f -> continuous_map X Tx Y Ty f -> homeomorphism X Tx Y Ty f! \\
\hline

Th~26.7 &
Th 26.7. The product of finitely many compact spaces is compact. &
\texttt{finite\_\allowbreak product\_\allowbreak compact} &
\Verb[breaklines=true,breakanywhere=true]!forall X Tx Y Ty:set, compact_space X Tx -> compact_space Y Ty -> compact_space (setprod X Y) (product_topology X Tx Y Ty)! \\
\hline

Lem~26.8 &
Lem 26.8 (The tube lemma). Consider the product space $X \times Y$, where $Y$ is compact. If $N$ is an open set of $X \times Y$ containing the slice $\{x_{0}\} \times Y$, then there is a neighborhood $W$ of $x_{0}$ in $X$ such that $W \times Y \subset N$. &
\texttt{tube\_\allowbreak lemma} &
\Verb[breaklines=true,breakanywhere=true]!forall X Tx Y Ty:set, topology_on X Tx -> topology_on Y Ty -> compact_space Y Ty -> forall x0 N:set, x0 :e X -> N :e product_topology X Tx Y Ty -> (setprod {x0} Y) c= N -> exists W:set, W :e Tx /\ x0 :e W /\ (setprod W Y) c= N! \\
\hline

Th~27.3 &
Th 27.3 (Heine--Borel theorem). A subspace $A$ of $\mathbb{R}$ is compact if and only if it is closed and bounded in $\mathbb{R}$. &
\texttt{Heine\_\allowbreak Borel\_\allowbreak closed\_\allowbreak bounded} &
\Verb[breaklines=true,breakanywhere=true]!forall A:set, A c= R -> compact_space A (subspace_topology R R_standard_topology A) <-> (closed_in R R_standard_topology A /\ bounded_set A)! \\
\hline

Th~28.1 &
Th 28.1. Compactness implies limit point compactness, but not conversely. &
\texttt{compact\_\allowbreak implies\_\allowbreak limit\_point\_\allowbreak compact} &
\Verb[breaklines=true,breakanywhere=true]!forall X Tx:set, compact_space X Tx -> limit_point_compact X Tx! \\
\hline

Th~30.3 &
Th 30.3. Suppose that $X$ has a countable basis. Then every open covering of $X$ contains a countable subcollection that covers $X$. &
\texttt{countable\_\allowbreak basis\_\allowbreak implies\_\allowbreak Lindelof} &
\Verb[breaklines=true,breakanywhere=true]!forall X Tx:set, second_countable_space X Tx -> Lindelof_space X Tx! \\
\hline

Th~30.4 &
Th 30.4. If $X$ has a countable basis, then $X$ contains a countable dense subset. &
\texttt{countable\_\allowbreak basis\_\allowbreak implies\_\allowbreak separable} &
\Verb[breaklines=true,breakanywhere=true]!forall X Tx:set, second_countable_space X Tx -> separable_space X Tx! \\
\hline

Th~32.1 &
Th 32.1. If $X$ is regular and has a countable basis, then $X$ is normal. &
\texttt{regular\_\allowbreak countable\_\allowbreak basis\_\allowbreak normal} &
\Verb[breaklines=true,breakanywhere=true]!forall X Tx:set, regular_space X Tx -> second_countable_space X Tx -> normal_space X Tx! \\
\hline

Th~32.3 &
Th 32.3. Every compact Hausdorff space is normal. &
\texttt{compact\_\allowbreak Hausdorff\_\allowbreak normal} &
\Verb[breaklines=true,breakanywhere=true]!forall X Tx:set, compact_space X Tx -> Hausdorff_space X Tx -> normal_space X Tx! \\
\hline

Th~33.1 &
Th 33.1 (Urysohn lemma). Let $X$ be a normal space; let $A$ and $B$ be disjoint closed subsets of $X$. Then there exists a continuous function $f:X\to[0,1]$ such that $f(A)=\{0\}$ and $f(B)=\{1\}$. &
\texttt{Urysohn\_\allowbreak lemma} &
\Verb[breaklines=true,breakanywhere=true]!forall X Tx A B:set, normal_space X Tx -> closed_in X Tx A -> closed_in X Tx B -> A :/\: B = Empty -> exists f:set, continuous_map X Tx unit_interval unit_interval_topology f /\ (forall a:set, a :e A -> apply_fun f a = 0) /\ (forall b:set, b :e B -> apply_fun f b = 1)! \\
\hline

Th~34.1 &
Th 34.1 (Urysohn metrization theorem). Every regular space with a countable basis is metrizable. &
\texttt{Urysohn\_\allowbreak metrization\_\allowbreak theorem} &
\Verb[breaklines=true,breakanywhere=true]!forall X Tx:set, regular_space X Tx -> second_countable_space X Tx -> metrizable X Tx! \\
\hline

Th~35.1 &
Th 35.1 (Tietze extension theorem). Let $X$ be a normal space and $A$ a closed subspace of $X$. Then every continuous map $f:A\to[-1,1]$ extends to a continuous map $F:X\to[-1,1]$. &
\texttt{Tietze\_\allowbreak extension\_\allowbreak interval} &
\Verb[breaklines=true,breakanywhere=true]!forall X Tx A f:set, normal_space X Tx -> closed_in X Tx A -> continuous_map A (subspace_topology X Tx A) closed_interval_minus1_1 closed_interval_minus1_1_topology f -> exists F:set, continuous_map X Tx closed_interval_minus1_1 closed_interval_minus1_1_topology F /\ forall a:set, a :e A -> apply_fun F a = apply_fun f a! \\
\hline

Th~41.1 &
Th 41.1. Every paracompact Hausdorff space is normal. &
\texttt{paracompact\_\allowbreak Hausdorff\_\allowbreak normal} &
\Verb[breaklines=true,breakanywhere=true]!forall X Tx:set, paracompact_space X Tx -> Hausdorff_space X Tx -> normal_space X Tx! \\
\hline

Th~41.2 &
Th 41.2. Every closed subspace of a paracompact space is paracompact. &
\texttt{closed\_\allowbreak subspace\_\allowbreak paracompact} &
\Verb[breaklines=true,breakanywhere=true]!forall X Tx Y:set, paracompact_space X Tx -> closed_in X Tx Y -> paracompact_space Y (subspace_topology X Tx Y)! \\
\hline

Lem~41.3 &
Lem 41.3 (Michael lemma). Let $X$ be regular. Then every open cover of $X$ has an open locally finite refinement if and only if it has an open $\sigma$-locally finite refinement. &
\texttt{Michael\_\allowbreak lemma\_\allowbreak 41\_\allowbreak 3} &
\Verb[breaklines=true,breakanywhere=true]!forall X Tx:set, regular_space X Tx -> ( (forall Ufam:set, open_cover X Tx Ufam -> exists Vfam:set, open_refinement X Tx Ufam Vfam /\ locally_finite_family X Tx Vfam) <-> (forall Ufam:set, open_cover X Tx Ufam -> exists Vfam:set, open_refinement X Tx Ufam Vfam /\ sigma_locally_finite_family X Tx Vfam) )! \\
\hline

Th~41.5 &
Th 41.5. Every regular Lindelöf space is paracompact. &
\texttt{regular\_\allowbreak Lindelof\_\allowbreak paracompact} &
\Verb[breaklines=true,breakanywhere=true]!forall X Tx:set, regular_space X Tx -> Lindelof_space X Tx -> paracompact_space X Tx! \\
\hline

Th~46.10 &
Th 46.10. Let $X$ be locally compact Hausdorff. Then the evaluation map $e:X\times C_k(X,Y)\to Y$ is continuous. &
\texttt{evaluation\_\allowbreak map\_\allowbreak continuous} &
\Verb[breaklines=true,breakanywhere=true]!forall X Tx Y Ty:set, locally_compact_space X Tx -> Hausdorff_space X Tx -> topology_on Y Ty -> continuous_map (setprod X (Ck X Tx Y Ty)) (product_topology X Tx (Ck X Tx Y Ty) (compact_open_topology X Tx Y Ty)) Y Ty (evaluation_map X Y)! \\
\hline

Th~22.2 &
Th 22.2. Let $p:X\to Y$ be a quotient map. If $A\subset X$ is saturated, then the restriction $p|_A:A\to p(A)$ is a quotient map. &
\texttt{restriction\_\allowbreak quotient\_\allowbreak saturated} &
\Verb[breaklines=true,breakanywhere=true]!forall X Tx Y Ty p A:set, quotient_map X Tx Y Ty p -> saturated_set X p A -> quotient_map A (subspace_topology X Tx A) (image_of A p) (subspace_topology Y Ty (image_of A p)) (restriction_map A p)! \\
\hline

Th~17.11 &
Th 17.11. Every simply ordered set is Hausdorff in the order topology. &
\texttt{simply\_\allowbreak ordered\_\allowbreak set\_\allowbreak Hausdorff} &
\Verb[breaklines=true,breakanywhere=true]!forall X Lt:set, simply_ordered_set X Lt -> Hausdorff_space X (order_topology X Lt)! \\
\hline

Lem~23.1 &
Lem 23.1. If $X$ admits a separation, then $X$ is not connected. &
\texttt{connected\_\allowbreak space\_\allowbreak no\_\allowbreak separation} &
\Verb[breaklines=true,breakanywhere=true]!forall X Tx:set, topology_on X Tx -> (exists A B:set, separation_of X Tx A B) -> ~ connected_space X Tx! \\
\hline

Th~23.4 &
Th 23.4. Let $A$ be a connected subspace of $X$. If $A \subset B \subset \overline{A}$, then $B$ is connected. &
\texttt{connected\_\allowbreak with\_\allowbreak limit\_\allowbreak points} &
\Verb[breaklines=true,breakanywhere=true]!forall X Tx A B:set, topology_on X Tx -> connected_space A (subspace_topology X Tx A) -> A c= B -> B c= closure_of X Tx A -> connected_space B (subspace_topology X Tx B)! \\
\hline

Th~24.3 &
Th 24.3 (Intermediate value theorem). Let $f:X\to Y$ be continuous, where $X$ is connected and ordered, and $Y$ has the order topology. Then $f(X)$ is an interval. &
\texttt{intermediate\_\allowbreak value\_\allowbreak theorem} &
\Verb[breaklines=true,breakanywhere=true]!forall X Tx Y f a b r:set, simply_ordered_set X -> topology_on X Tx -> topology_on Y (order_topology Y) -> continuous_map X Tx Y (order_topology Y) f -> between (apply_fun f a) r (apply_fun f b) -> exists c:set, c :e X /\ apply_fun f c = r! \\
\hline

Th~25.1 &
Th 25.1. The components of a space $X$ form a partition of $X$ into connected subspaces. &
\texttt{components\_\allowbreak partition\_\allowbreak space} &
\Verb[breaklines=true,breakanywhere=true]!forall X Tx:set, topology_on X Tx -> partition_of X (components_of X Tx)! \\
\hline

Th~25.5 &
Th 25.5. If $X$ is locally connected, then the components of open sets of $X$ are open. &
\texttt{components\_\allowbreak are\_\allowbreak open\_in\_locally\_\allowbreak connected} &
\Verb[breaklines=true,breakanywhere=true]!forall X Tx:set, locally_connected_space X Tx -> forall C:set, C :e components_of X Tx -> open_in X Tx C! \\
\hline

Lem~26.1 &
Lem 26.1. A subspace $Y$ of $X$ is compact iff every open cover of $Y$ by sets open in $X$ has a finite subcover. &
\texttt{compact\_\allowbreak space\_\allowbreak subcover\_\allowbreak property} &
\Verb[breaklines=true,breakanywhere=true]!forall X Tx Y:set, topology_on X Tx -> Y c= X -> (compact_space Y (subspace_topology X Tx Y) <-> (forall Cfam:set, (forall U:set, U :e Cfam -> U :e Tx) -> Y c= Union Cfam -> exists F:set, F c= Cfam /\ finite_set F /\ Y c= Union F))! \\
\hline

Lem~26.4 &
Lem 26.4. If $Y$ is a compact subspace of Hausdorff $X$ and $x_0\notin Y$, then $x_0$ and $Y$ have disjoint neighborhoods. &
\texttt{Hausdorff\_\allowbreak separate\_\allowbreak point\_compact\_\allowbreak set} &
\Verb[breaklines=true,breakanywhere=true]!forall X Tx Y x0:set, Hausdorff_space X Tx -> compact_space Y (subspace_topology X Tx Y) -> x0 :e X -> x0 :e: Y -> exists U V:set, U :e Tx /\ V :e Tx /\ x0 :e U /\ Y c= V /\ U :/\: V = Empty! \\
\hline

Lem~27.5 &
Lem 27.5 (Lebesgue number lemma). Every open cover of a compact metric space has a Lebesgue number. &
\texttt{Lebesgue\_\allowbreak number\_\allowbreak lemma} &
\Verb[breaklines=true,breakanywhere=true]!forall X d:set, metric_space X d -> compact_space X (metric_topology X d) -> forall Cfam:set, open_cover X (metric_topology X d) Cfam -> exists e:set, Lebesgue_number X d Cfam e! \\
\hline

Th~27.6 &
Th 27.6. A continuous function from a compact space to a metric space is uniformly continuous. &
\texttt{uniform\_\allowbreak continuity\_\allowbreak on\_\allowbreak compact} &
\Verb[breaklines=true,breakanywhere=true]!forall X Tx Y d f:set, compact_space X Tx -> metric_space Y d -> continuous_map X Tx Y (metric_topology Y d) f -> uniformly_continuous X Y d f! \\
\hline

Th~29.1 &
Th 29.1. A locally compact Hausdorff space has a one-point compactification. &
\texttt{one\_\allowbreak point\_\allowbreak compactification\_\allowbreak exists} &
\Verb[breaklines=true,breakanywhere=true]!forall X Tx:set, locally_compact_Hausdorff X Tx -> exists Y Ty p:set, one_point_compactification_of X Tx Y Ty p! \\
\hline

Th~30.2 &
Th 30.2. Subspaces and countable products of first-countable spaces are first-countable. &
\texttt{countability\_\allowbreak axioms\_\allowbreak subspace\_\allowbreak product} &
\Verb[breaklines=true,breakanywhere=true]!forall X Tx:set, first_countable_space X Tx -> (forall Y:set, Y c= X -> first_countable_space Y (subspace_topology X Tx Y)) /\ (forall Fam:set, (forall i:set, i :e index_set Fam -> first_countable_space (Fam[i]) (space_family_topology Fam[i])) -> first_countable_space (product_space Fam) (product_topology_full Fam))! \\
\hline

Lem~31.1 &
Lem 31.1. In a $T_1$ space, singletons are closed. &
\texttt{T1\_\allowbreak space\_\allowbreak one\_point\_sets\_\allowbreak closed} &
\Verb[breaklines=true,breakanywhere=true]!forall X Tx:set, T1_space X Tx -> forall x:set, x :e X -> closed_in X Tx {x}! \\
\hline

Th~48.1 &
Th 48.1. A space $X$ is a Baire space iff every countable intersection of dense open sets is dense. &
\texttt{Baire\_\allowbreak space\_\allowbreak dense\_\allowbreak Gdelta} &
\Verb[breaklines=true,breakanywhere=true]!forall X Tx:set, Baire_space X Tx <-> (forall Ufam:set, countable_set Ufam -> (forall U:set, U :e Ufam -> open_in X Tx U /\ dense_in X Tx U) -> dense_in X Tx (Inter Ufam))! \\
\hline

\end{longtable}

\end{document}